\documentclass[a4paper,12pt]{article}
\usepackage{amsmath}
\usepackage{amssymb}
\usepackage{amsmath,amsthm,amssymb,amscd}
\usepackage{latexsym}
\usepackage{indentfirst}
\usepackage{subfigure}
\usepackage{graphicx}
\usepackage[top=1in,bottom=1in,left=1.25in,right=1.25in]{geometry}

\makeatother
\date{}
\theoremstyle{plain}

\begin{document}

\bibliographystyle{unsrt}
\bibliographystyle{plain}

\title{
  \bf Limit properties of periodic one dimensional hopping model
}

\author{{Yunxin Zhang}\thanks{
School of Mathematical Sciences, Fudan University,  Shanghai 200433,
China}
\thanks{Centre for Computational Systems Biology, Fudan University   (E-Mail: xyz@fudan.edu.cn)}
}

\maketitle \baselineskip=6mm

\baselineskip=22pt

\begin{abstract}
Periodic one dimensional hopping model is useful to study the motion
of microscopic particles, which lie in thermal noise environment.
The mean velocity $V_N$ and diffusion constant $D_N$ of this model
have been obtained by Bernard Derrida [J. Stat. Phys. 31 (1983)
433]. In this research, we will give the limits $V_D$ and $D_D$ of
$V_N$ and $D_N$ as the number $N$ of mechanochemical sates in one
period tends to infinity by formal calculation. It is well known
that the stochastic motion of microscopic particles also can be
described by overdamped Langevin dynamics and Fokker-Planck
equation. Up to now, the corresponding formulations of mean velocity
and effective diffusion coefficient, $V_L$ and $D_L$ in the
framework of Langevin dynamics and $V_P, D_P$ in the framework of
Fokker-Planck equation, have also been known. In this research, we
will find that the formulations $V_D$ and $V_L, V_P$ are
theoretically equivalent, and numerical comparison indicates that
$D_D, D_L$, and $D_P$ are almost the same. Through the discussion in
this research, we also can know more about the relationship between
the one dimensional hopping model and Fokker-Planck equation.

\vspace{2em} \noindent \textit{PACS}: 87.16.Nn, 87.16.A-, 82.39.-k,
05.40.Jc

\vspace{2em} \noindent \textit{Keywords}: diffusion coefficient;
Langevin equation; Fokker-Planck equation; hopping model
\end{abstract}

\section{Introduction}
Many physical \cite{Alexander1981, Alexander19811} and biochemical
phenomena, especially for the motion of motor protein
\cite{Carter2005, Zhang20081, Kolomeisky2000, Lattanzi2001}, can
be described by the periodic one dimensional hopping model. In
this model, the particle jumps along a periodical linear track
from one binding site to next one through the sequence of $N$
mechanochemical states \cite{Fisher1999, Zhang20091}. The particle
in state $j$ can jump forward to state $j+1$ with the rate $u_j$,
or jump backward to state $j-1$ with the rate $w_{j}$. After
moving $N$ sites forward the particle comes to the same
mechanochemical state but shifted by a step size distance $L$ (for
motor protein kinesin $L=8.2$nm). The dynamics of particle motion
is described by the standard rate equations of occupation
probabilities $p_j(t)$
\begin{equation}\label{eq1}
\begin{aligned}
\frac{\partial p_j(t)}{\partial
t}=&u_{j-1}p_{j-1}(t)+w_{j+1}p_{j+1}(t)-[u_j+w_j]p_j(t)\cr &\quad
0\le j\le N
\end{aligned}
\end{equation}
where
\begin{equation}\label{eq2}
p_{lN+j}(t)=p_j(t)\quad u_{lN+j}=u_j\quad w_{lN+j}=w_j\quad l\in
\mathcal{Z}
\end{equation}
At steady state,
\begin{equation}\label{eq3}
u_{j-1}p_{j-1}+w_{j+1}p_{j+1}-[u_j+w_j]p_j=0\quad 0\le j\le N
\end{equation}
It's solution is
\begin{equation}\label{eq4}
p_j=\frac{r_j}{R_N}
\end{equation}
where
\begin{equation}\label{eq5}
r_j=\frac{1}{u_j}\left[1+\sum_{k=1}^{N-1}\prod_{i=j+1}^{j+k}\frac{w_i}{u_i}\right]\qquad
R_N=\sum_{j=0}^{N-1}r_j
\end{equation}
This model has been extensively studied \cite{Webman1981,
Machta1982} and its mean velocity $V_N$ and diffusion constant $D_N$
has been obtained explicitly \cite{Derrida1983}.
\begin{equation}\label{eq6}
V_N=\frac{L\left[1-\prod\limits_{j=0}^{N-1}\frac{w_i}{u_i}\right]}{R_N}\qquad
D_N=\frac{L}{N}\left[\frac{LU_N+VS_N}{R_N^2}-\frac{(N+2)V}{2}\right]
\end{equation}
where
$$
\begin{aligned}
S_N=\sum_{j=0}^{N-1}s_j\sum_{k=0}^{N-1}(k+1)r_{k+j+1}\quad
U_N=\sum_{j=0}^{N-1}u_jr_js_j\quad
s_j=\frac{1}{u_j}\left[1+\sum_{k=1}^{N-1}\prod_{i=j-1}^{j-k}\frac{w_{i+1}}{u_i}\right]
\end{aligned}
$$

Besides one dimensional hopping model, the stochastic motion of
particle also can be modelled by Langevin dynamics
\cite{Reimann2008, Reimann2009} and Fokker-Planck equation
\cite{Howard2001, Zhang20091, Zhang2009, Risken1989}. Intuitively,
the Langevin dynamics and Fokker-Planck equation can be regarded
as infinite mechanochemical states case of periodic one
dimensional hopping model. In the framework of Langevin dynamics,
the particle position is governed by the following Langevin
equation
\begin{equation}\label{7}
\xi \frac{d x(t)}{d t}=-\frac{\partial \phi(x)}{\partial
x}+\sqrt{2k_BT\xi}f_B(t)
\end{equation}
where $\xi$ is viscous friction coefficient, $k_B$ is Boltzmann's
constant. $\phi(x)=\Phi(x)-F_{ext}x$, $F_{ext}$ is external load,
$\Phi$ is a (tilted) periodic potential with period $L$. $T$ is
absolute temperature and $f_B(t)$ is Gaussian white noise. Based on
the relation between the effective diffusion coefficient and the
first two moments of the mean first passage time (MFPT), the mean
velocity and effective diffusion coefficient can be expressed in
quadratures (\cite{Reimann2002} \cite{Reimann2001})
\begin{equation}\label{8}
V_L=\frac{1-e^{\frac{\Delta\phi}{k_BT}}}{\int_0^LdxI_+(x)}L\qquad
{D}_{L}=\frac{\int_0^LdxI_+^2(x)I_-(x)}{\left[\int_0^LdxI_+(x)\right]^3}DL^2
\end{equation}
where
\begin{equation}\label{9}
I_{\pm}=\mp\frac{e^{\pm\phi(x)/K_BT}}{D}\int_x^{x\mp
L}dye^{\mp\phi(y)/K_BT}
\end{equation}
$\Delta\phi\triangleq\phi(L)-\phi(0)$ and $D$ is the free diffusion
constant which satisfies the Einstein relation $\xi D=k_BT$.

At the same time, the stochastic motion of particle can be
modelled by the Fokker-Planck equation \cite{Howard2001,
Zhang2009, Risken1989}
\begin{equation}\label{eq10}
\frac{\partial \rho}{\partial t}=\frac{\partial}{\partial
x}\left[\frac{\rho}{\xi}\frac{\partial \phi}{\partial
x}+D\frac{\partial \rho}{\partial x}\right]\qquad 0\le x\le L
\end{equation}
where $\rho(x, t)$ is the probability density for finding particles
at position $x$ and time $t$. At steady state,
\begin{equation}\label{eq11}
\frac{\partial}{\partial x}\left[\frac{\rho}{\xi}\frac{\partial
\phi}{\partial x}+D\frac{\partial \rho}{\partial x}\right]=0\qquad
0\le x\le L
\end{equation}
It's solution is (see \cite{Howard2001, Reimann2002,
Evstigneev2008})
\begin{equation}\label{eq12}
\rho(x)=\frac{J\exp{\left(-\frac{\phi(x)}{k_BT}\right)}}{D\left[1-\exp{\left(\frac{\Delta\phi}{k_BT}\right)}\right]}
\left(\int_x^{x+L}\exp{\left(\frac{\phi(y)}{k_BT}\right)}dy\right)
\end{equation}
where
\begin{equation}\label{eq13}
J=\frac{D\left[1-\exp{\left(\frac{\Delta\phi}{k_BT}\right)}\right]}{\int_0^L\exp{\left(-\frac{\phi(x)}{k_BT}\right)}
\left(\int_x^{x+L}\exp{\left(\frac{\phi(y)}{k_BT}\right)}dy\right)dx}
\end{equation}
is the probability flux. So the mean velocity of particles is
\begin{equation}\label{eq14}
V_F=LJ=\frac{DL\left[1-\exp{\left(\frac{\Delta\phi}{k_BT}\right)}\right]}{\int_0^L\exp{\left(-\frac{\phi(x)}{k_BT}\right)}
\left(\int_x^{x+L}\exp{\left(\frac{\phi(y)}{k_BT}\right)}dy\right)dx}
\end{equation}
In \cite{Zhang2008} we have known that the formulation of the
effective diffusion coefficient is
\begin{equation}\label{eq15}
\begin{aligned}
D_F =&J\int_0^L\left[(L\rho(x)-1)\int_0^x(L\rho(z)-1)dz\right]dx\cr
&-\frac{JL}{\exp{\left(\frac{\triangle\phi}{k_BT}\right)}-1}
\int_0^L\exp{\left(-\frac{\phi(x)}{k_BT}\right)}\left[\int_x^{x+L}\rho^2(z)\exp{\left(\frac{\phi(z)}{k_BT}\right)}dz\right]d\,x
\end{aligned}
\end{equation}

As we have mentioned, the continuous models (Langevin equation and
Fokker-Planck equation) can be regarded as infinite mechanochemical
states cases of the discrete model, i.e. the periodic one
dimensional hopping model. Therefore, the mean velocity and
effective diffusion coefficient of continuous cases can be obtained
directly by calculating the limits of which in discrete case
(\ref{eq6}). In the following, we will obtain the limits $V_D$ and
$D_D$ of $V_N$ and $D_N$ by formal calculation. Theoretically, $V_D$
and $V_L, V_P$ are equivalent. Though $D_D$ and $D_L, D_P$ are not
theoretically equivalent, since $D_N, D_L$ and $D_P$ are all
obtained under some assumptions, numerical results indicate that the
differences among them are very small.

\section{The limit of velocity of the hopping model}
Firstly, we consider the large mechanochemical state $N$ limit of
the velocity of one dimensional hopping model.

Due to the detailed balance
\begin{equation}\label{eq16}
\frac{w_{i+1}}{u_i}=\frac{\exp{\left(\frac{\phi(x_{i+1})}{k_BT}\right)}}{\exp{\left(\frac{\phi(x_{i})}{k_BT}\right)}}
=\exp{\left(\frac{\triangle\phi_i}{k_BT}\right)}
\end{equation}
where $x_i=\frac{iL}{N}$, $\Delta \phi_i=\phi(x_{i+1})-\phi(x_{i})$,
one can know that
\begin{equation}\label{eq17}
\begin{aligned}
r_j=&\frac{1}{u_j}\left[1+\sum_{k=1}^{N-1}\prod_{i=j+1}^{j+k}\frac{w_i}{u_i}\right]\cr
=&\frac{1}{u_{j}}+\sum_{k=1}^{N-1}\frac{1}{u_{j+k}}\prod_{i=j+1}^{j+k}\frac{w_i}{u_{i-1}}\cr
=&\frac{1}{u_{j}}+\sum_{k=1}^{N-1}\frac{1}{u_{j+k}}\prod_{i=j+1}^{j+k}\exp{\left(\frac{\Delta
\phi_{i-1}}{k_BT}\right)}\cr
=&\frac{1}{u_{j}}+\sum_{k=1}^{N-1}\frac{1}{u_{j+k}}\exp{\left(\frac{\phi(x_{j+k})-\phi(x_{j})}{k_BT}\right)}\cr
=&\sum_{k=0}^{N-1}\frac{1}{u_{j+k}}\exp{\left(\frac{\phi(x_{j+k})-\phi(x_{j})}{k_BT}\right)}\cr
\end{aligned}
\end{equation}
In the large $N$ limit, $u_j\approx
\overline{D}_N\left(\frac{N}{L}\right)^2=\overline{D}_N/(\Delta
x)^2$ (for detailed discussion, see \cite{Fricks2006,
Lattanzi2001}). Therefore
\begin{equation}\label{eq18}
\begin{aligned}
r(x_j)\approx
&\frac{N}{L}\int_0^L\frac{1}{u(x_{j}+z)}\exp{\left(\frac{\phi(x_{j}+z)-\phi(x_{j})}{k_BT}\right)}dz\cr
=&\frac{N}{L}\int_{x_j}^{x_j+L}\frac{1}{u(z)}\exp{\left(\frac{\phi(z)-\phi(x_{j})}{k_BT}\right)}dz\cr
\approx&\frac{1}{\overline{D}_N\left(\frac{N}{L}\right)}\int_{x_j}^{x_j+L}\exp{\left(\frac{\phi(z)-\phi(x_{j})}{k_BT}\right)}dz
\end{aligned}
\end{equation}
consequently
\begin{equation}\label{eq19}
\begin{aligned}
R_N=&\sum_{j=0}^{N-1}r_j\approx\sum_{j=0}^{N-1}\frac{1}{\overline{D}_N\left(\frac{N}{L}\right)}\int_{x_j}^{x_j+L}\exp{\left(\frac{\phi(z)-\phi(x_{j})}{k_BT}\right)}dz\cr
\approx&\frac{1}{\overline{D}_N}\int_0^L\left[\int_{x}^{x+L}\exp{\left(\frac{\phi(z)-\phi(x_{j})}{k_BT}\right)}dz\right]dx\cr
\end{aligned}
\end{equation}
Thanks to the periodicity of the transition rates $u_j$ and $w_j$
(\ref{eq2}),
\begin{equation}\label{eq20}
\begin{aligned}
\prod\limits_{j=0}^{N-1}\frac{w_i}{u_i}=&\prod\limits_{j=0}^{N-1}\frac{w_{i+1}}{u_i}=\exp{\left(\frac{\phi(x_{N})-\phi(x_{0})}{k_BT}\right)}=\exp{\left(\frac{\Delta
\phi}{k_BT}\right)}
\end{aligned}
\end{equation}
From (\ref{eq2}) (\ref{eq19}) (\ref{eq20}), it is easy to obtain
\begin{equation}\label{eq21}
V_D:=\lim_{N\to\infty}V_N=\frac{\overline{D}L\left[1-\exp{\left(\frac{\Delta
\phi}{k_BT}\right)}\right]}
{\int_0^L\left[\int_{x}^{x+L}\exp{\left(\frac{\phi(z)-\phi(x)}{k_BT}\right)}dz\right]dx}
\end{equation}
where $\overline{D}=\lim_{N\to\infty}\overline{D}_N$. One can easily
verify that, if $\overline{D}=D$ then $V_D=V_L=V_F$ (see (\ref{8})
(\ref{eq14})). In the following, we always assume that
$\overline{D}=D$.

\section{The limit of effective diffusion coefficient of the hopping model}
Using the similar discussion as in the above section, it can be
easily verified that
\begin{equation}\label{eq22}
\begin{aligned}
s_j=&\frac{1}{u_j}\left[1+\sum_{k=1}^{N-1}\prod_{i=j-1}^{j-k}\frac{w_{i+1}}{u_i}\right]\cr
=&\frac{1}{u_{j}}\left[1+\sum_{k=1}^{N-1}\prod_{i=j-1}^{j-k}\exp{\left(\frac{\Delta
\phi_{i}}{k_BT}\right)}\right]\cr
=&\frac{1}{u_{j}}\left[1+\sum_{k=1}^{N-1}\exp{\left(\frac{\phi(x_{j})-\phi(x_{j-k})}{k_BT}\right)}\right]\cr
=&\frac{1}{u_{j}}\sum_{k=0}^{N-1}\exp{\left(\frac{\phi(x_{j})-\phi(x_{j-k})}{k_BT}\right)}\cr
\end{aligned}
\end{equation}
In the large biochemical state number $N$ limit,
\begin{equation}\label{eq23}
\begin{aligned}
s(x_j)\approx
&\frac{N}{L}\frac{1}{u(x_{j})}\int_0^L\exp{\left(\frac{\phi(x_{j})-\phi(x_{j}-z)}{k_BT}\right)}dz\cr
=&\frac{N}{L}\frac{1}{u(x_{j})}\int_{x_j-L}^{x_j}\exp{\left(\frac{\phi(x_{j})-\phi(z)}{k_BT}\right)}dz\cr
\approx&\frac{1}{\overline{D}_N\left(\frac{N}{L}\right)}\int_{x_j-L}^{x_j}\exp{\left(\frac{\phi(x_{j})-\phi(z)}{k_BT}\right)}dz
\end{aligned}
\end{equation}
For the sake of the simplicity, we define
\begin{equation}\label{eq24}
\begin{array}{l}
f(x)=\int_{x}^{x+L}\exp{\left(\frac{\phi(z)-\phi(x)}{k_BT}\right)}dz\qquad
g(x)=\int_{x-L}^{x}\exp{\left(\frac{\phi(x)-\phi(z)}{k_BT}\right)}dz
\end{array}
\end{equation}
It can be readily verified that
\begin{equation}\label{eq25}
\begin{aligned}
&R_N\approx\frac{1}{\overline{D}_N}\int_0^Lf(x)dx\qquad
r(x_j)\approx\frac{f(x_j)}{\overline{D}_N\left(\frac{N}{L}\right)}\qquad
s(x_j)\approx\frac{g(x_j)}{\overline{D}_N\left(\frac{N}{L}\right)}
\end{aligned}
\end{equation}
\begin{equation}\label{eq26}
\begin{aligned}
U_N=&\sum_{j=0}^{N-1}u_jr_js_j
\approx\sum_{j=0}^{N-1}\overline{D}_N\left(\frac{N}{L}\right)^2
\left[\frac{f(x_j)}{\overline{D}_N\left(\frac{N}{L}\right)}\right]
\left[\frac{g(x_j)}{\overline{D}_N\left(\frac{N}{L}\right)}\right]\cr
=&\sum_{j=0}^{N-1}\frac{1}{\overline{D}_N}f(x_j)g(x_j)
=\frac{N}{L\overline{D}_N}\int_{0}^{L}f(x)g(x)dx
\end{aligned}
\end{equation}
Moreover
\begin{equation}\label{eq27}
\begin{aligned}
&\sum_{k=0}^{N-1}(k+1)r_{k+j+1}\approx
\sum_{k=0}^{N-1}(k+1)f(x_{k+j+1})\frac{1}{\overline{D}_N\left(\frac{N}{L}\right)}\cr
=&\sum_{k=0}^{N-1}x_{k+1}f(x_j+x_{k+1})\frac{1}{\overline{D}_N}
\approx\frac{N}{L\overline{D}_N}\int_{0}^Lzf(x_j+z)dz
\end{aligned}
\end{equation}
so
\begin{equation}\label{eq28}
\begin{aligned}
S_N=&\sum_{j=0}^{N-1}s_j\sum_{k=0}^{N-1}(k+1)r_{k+j+1}\cr
\approx&\sum_{j=0}^{N-1}\left[\frac{g(x_j)}{\overline{D}_N\left(\frac{N}{L}\right)}\right]
\left[\frac{N}{L\overline{D}_N}\int_{0}^Lzf(x_j+z)dz\right]\cr
=&\frac{1}{\overline{D}_N^2}\sum_{j=0}^{N-1}g(x_j)\int_{0}^Lzf(x_j+z)dz\cr
\approx&\frac{N}{L\overline{D}_N^2}\int_0^L\left[g(x)\int_{0}^Lzf(x+z)dz\right]dx
\end{aligned}
\end{equation}
Substituting (\ref{eq21}) (\ref{eq25}) (\ref{eq26}) (\ref{eq28})
into (\ref{eq6}), one obtains
\begin{equation}\label{eq29}
\begin{aligned}
D_N=&\frac{L}{N}\left[\frac{LU_N+VS_N}{R_N^2}-\frac{(N+2)V}{2}\right]\cr
=&\frac{L}{N}\left[\frac{L\left[1-\exp{\left(\frac{\Delta
\phi}{k_BT}\right)}\right]
\left[\frac{N}{L\overline{D}_N^2}\int_0^L\left[g(x)\int_{0}^Lzf(x+z)dz\right]dx\right]}
{\left[\frac{1}{\overline{D}_N}\int_0^Lf(x)dx\right]^3}\right.\cr
&+\frac{L\left[\frac{N}{L\overline{D}_N}
\int_{0}^{L}f(x)g(x)dx\right]}{\left[\frac{1}{\overline{D}_N}\int_0^Lf(x)dx\right]^2}
\left.-\frac{(N+2)\overline{D}_NL\left[1-\exp{\left(\frac{\Delta
\phi}{k_BT}\right)}\right]}{2\int_0^Lf(x)dx}\right]\cr
=&L\overline{D}_N\left[\frac{\left[1-\exp{\left(\frac{\Delta
\phi}{k_BT}\right)}\right]
\left[\int_0^L\left[g(x)\int_{0}^Lzf(x+z)dz\right]dx\right]}
{\left[\int_0^Lf(x)dx\right]^3}\right.\cr &\left.+\frac{
\int_{0}^{L}f(x)g(x)dx}{\left[\int_0^Lf(x)dx\right]^2}
-\frac{(N+2)L\left[1-\exp{\left(\frac{\Delta
\phi}{k_BT}\right)}\right]}{2N\int_0^Lf(x)dx}\right]
\end{aligned}
\end{equation}
hence
\begin{equation}\label{eq30}
\begin{aligned}
\lim_{N\to\infty}D_N=&L\overline{D}\left[\frac{\left[1-\exp{\left(\frac{\Delta
\phi}{k_BT}\right)}\right]
\left[\int_0^L\left[g(x)\int_{0}^Lzf(x+z)dz\right]dx\right]}
{\left[\int_0^Lf(x)dx\right]^3}\right.\cr &+\frac{
\int_{0}^{L}f(x)g(x)dx}{\left[\int_0^Lf(x)dx\right]^2}\left.-\frac{L\left[1-\exp{\left(\frac{\Delta
\phi}{k_BT}\right)}\right]}{2\int_0^Lf(x)dx}\right]
\end{aligned}
\end{equation}
From (\ref{eq12}, \ref{eq13}) and (\ref{eq24}), one can know that
\begin{equation}\label{eq31}
\begin{aligned}
&f(x)=\frac{D\left[1-\exp{\left(\frac{\Delta
\phi}{k_BT}\right)}\right]}{J}\rho(x)\cr
&\int_0^Lf(x)dx=\frac{D\left[1-\exp{\left(\frac{\Delta
\phi}{k_BT}\right)}\right]}{J}
\end{aligned}
\end{equation}
Substituting (\ref{eq31}) into (\ref{eq30}), we finally get the
limit of the effective diffusion coefficient
\begin{equation}\label{eq32}
\begin{aligned}
D_{D}:=&\lim_{N\to\infty}D_N\cr
=&\frac{JL\int_0^L\rho(x)g(x)dx}{\left[1-\exp{\left(\frac{\Delta
\phi}{k_BT}\right)}\right]}+\frac{J^2L\int_0^L\left[g(x)\int_{0}^Lz\rho(x+z)dz\right]dx}{D\left[1-\exp{\left(\frac{\Delta
\phi}{k_BT}\right)}\right]}-\frac12JL^2\cr
=&\frac{V\int_0^L\rho(x)g(x)dx}{\left[1-\exp{\left(\frac{\Delta
\phi}{k_BT}\right)}\right]}+\frac{JV\int_0^L\left[g(x)\int_{0}^Lz\rho(x+z)dz\right]dx}{D\left[1-\exp{\left(\frac{\Delta
\phi}{k_BT}\right)}\right]}-\frac12VL
\end{aligned}
\end{equation}
By the way, the limit of the randomness parameter $
RAND_N=\frac{2D_N}{V_NL}$ is
\begin{equation}\label{eq33}
\begin{aligned}
\lim_{N\to\infty}RAND_N=&\frac{2\int_0^L\rho(x)g(x)dx}{L\left[1-\exp{\left(\frac{\Delta
\phi}{k_BT}\right)}\right]}+\frac{2J\int_0^L\left[g(x)\int_{0}^Lz\rho(x+z)dz\right]dx}{DL\left[1-\exp{\left(\frac{\Delta
\phi}{k_BT}\right)}\right]}-1
\end{aligned}
\end{equation}

\section{Numerical results and discussion}
As we have known that the limit $V_D$ of velocity of the hopping
model is the same as $V_L$ and $V_F$. However, it is easy to
understand that $D_D$ would be different from $D_L, D_F$ because
they are obtained under different assumptions.

\begin{figure}
  \includegraphics[width=200pt]{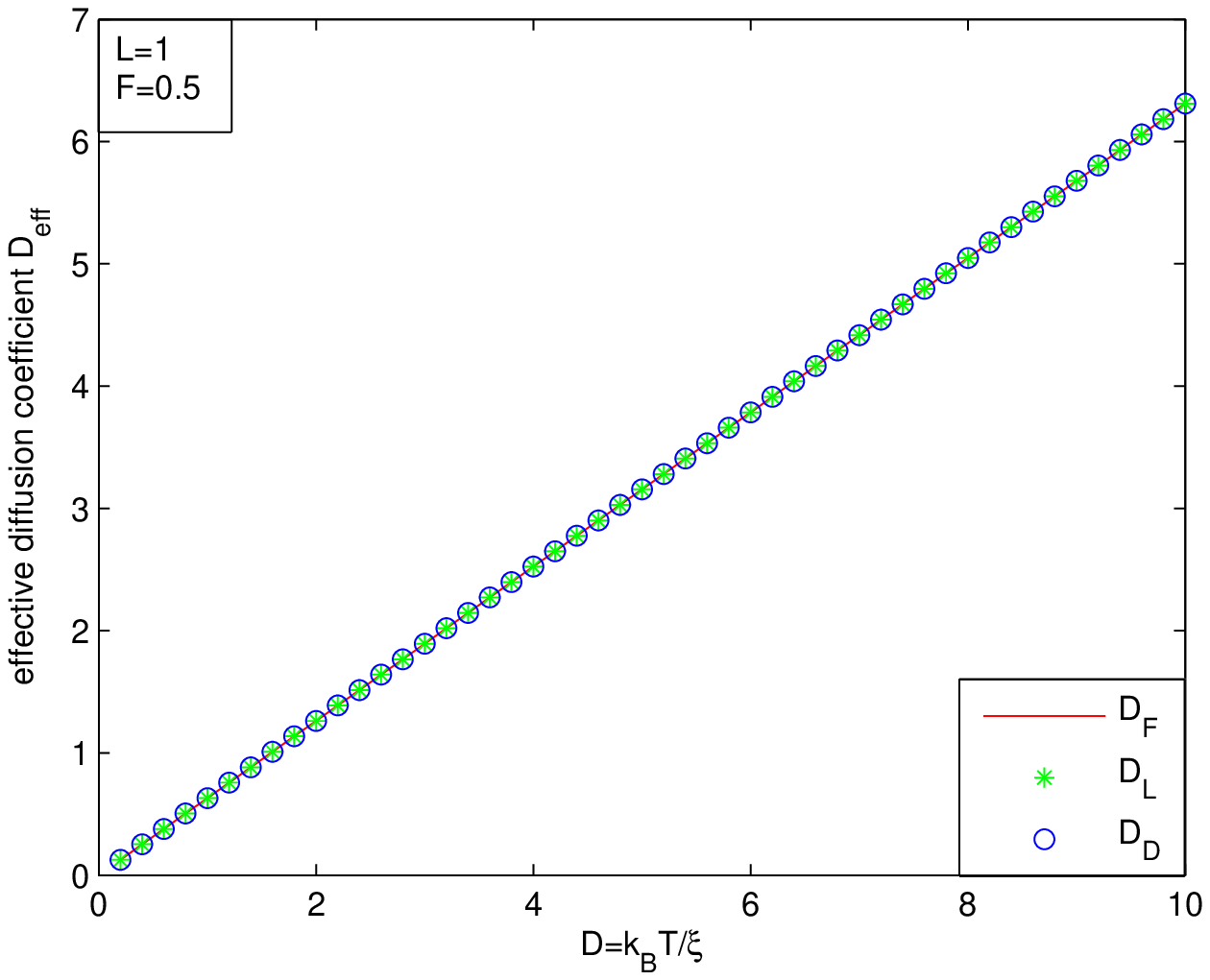} \includegraphics[width=200pt]{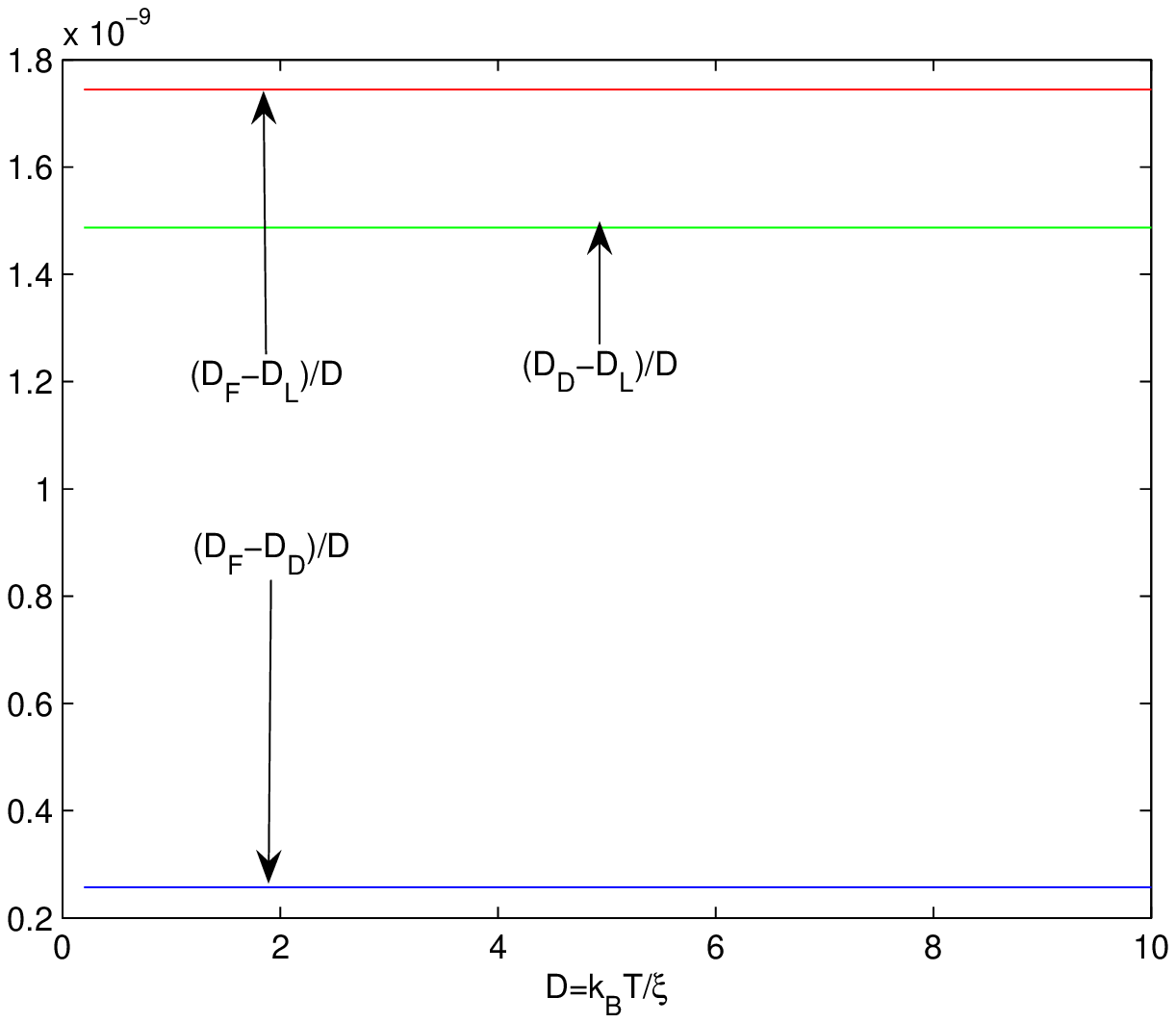}\\
  \caption{{\bf Left:} figures of effective diffusion coefficient
as a function of free diffusion coefficient $D:=k_BT/\xi$, where the
red line is obtained by the formulation (\ref{eq15}), the green
stars are obtained  by the formulation (\ref{8}) and the blue
circles are obtained by the formulation (\ref{eq32}). {\bf Right: }
the difference between the formulations (\ref{eq32}) (\ref{8}) and
(\ref{eq15}), where $D_F$ denote the results of formulation
(\ref{eq15}), $D_L$ denote the results of the formulation (\ref{8})
and $D_D$ denote the results of the formulation (\ref{eq32}). In the
simulation, the potential $\Phi(x)=U_0 sin(2\pi x/L))-F_{ext}x$ and
$\xi=U_0=L=1$ (see \cite{Reimann2001}), $F_{ext}=0.5$.}
\label{Figure1}
\end{figure}

\begin{figure}
  \includegraphics[width=200pt]{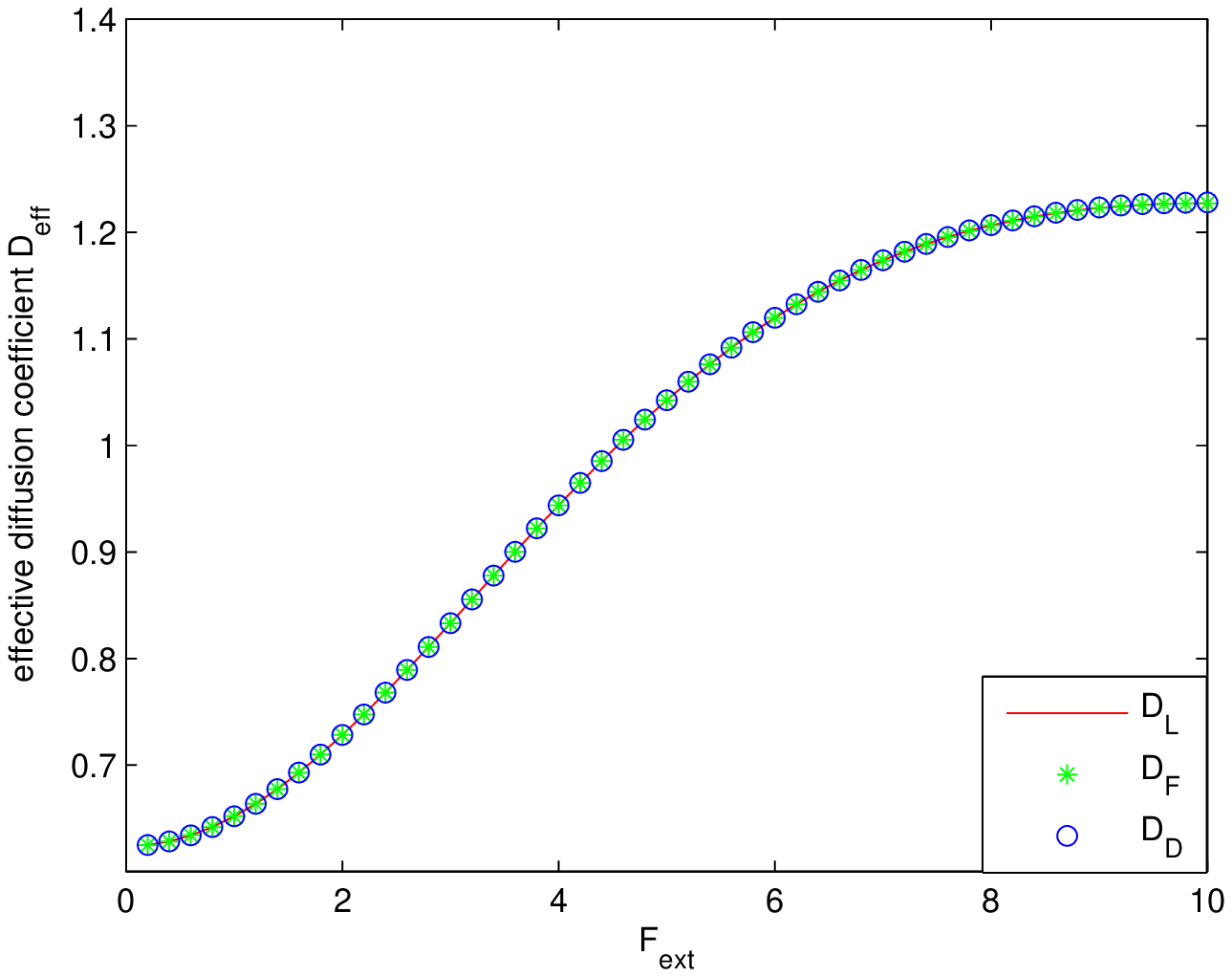}\includegraphics[width=200pt]{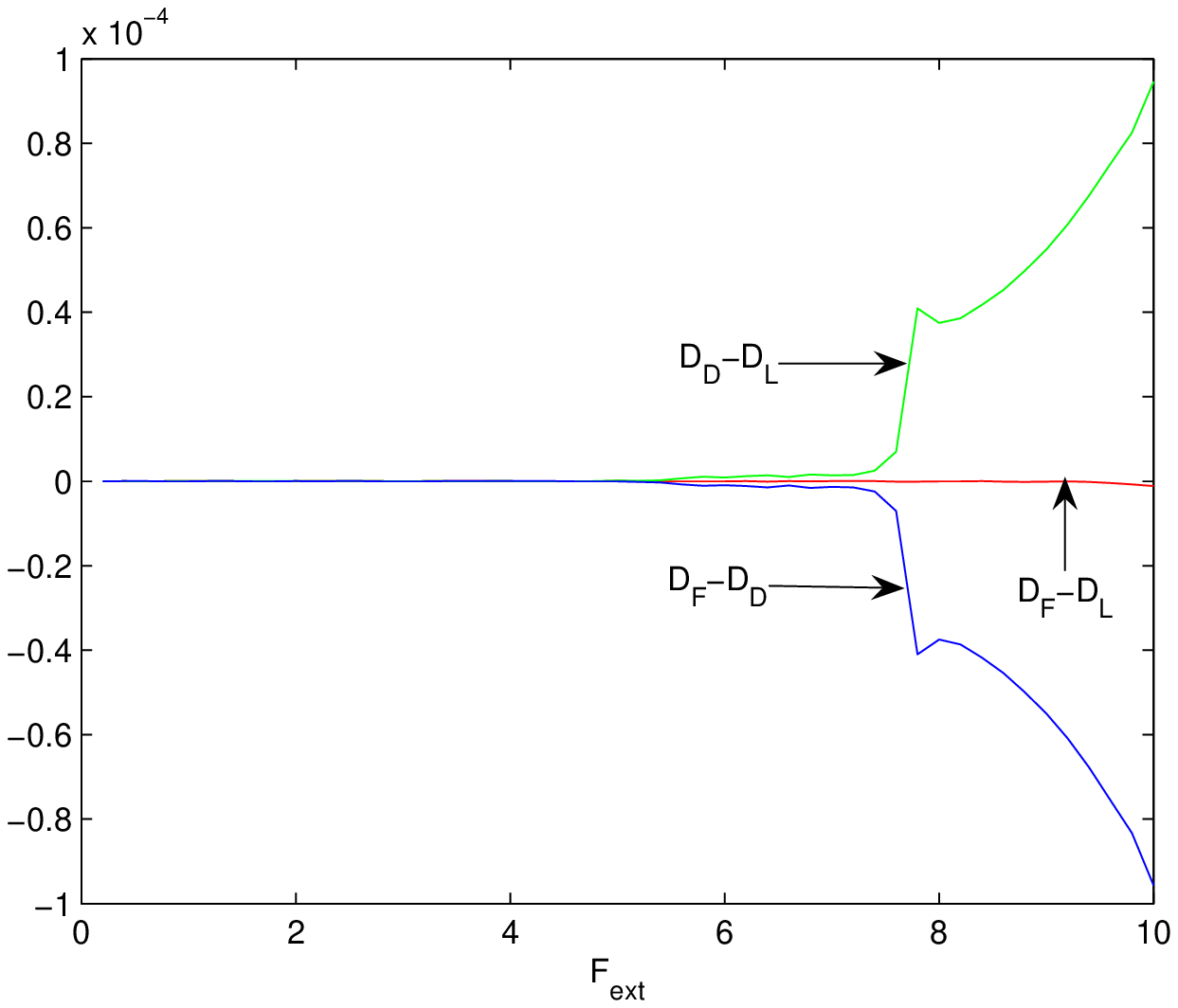}\\
   \caption{{\bf Left:} the relationship between the effective diffusion coefficient
 and the external force $F_{ext}$.
 {\bf Right: } the differences between the formulations
(\ref{8}) (\ref{eq15}) and (\ref{eq32}). In the simulation, the
potential $\Phi(x)=U_0 sin(2\pi x/L))-F_{ext}x$ and $\xi=U_0=L=1$
(see \cite{Reimann2001}).} \label{Figure2}
\end{figure}

To indicate the accuracy of the formulation (\ref{eq32}), we present
the numerical results of the formulations (\ref{8}) (\ref{eq15}) and
(\ref{eq32}) in Figure \ref{Figure1} and \ref{Figure2}. From the
numerical results we can find that the formulation (\ref{eq32}) is
accurate enough, especially for the cases with small external force
$F_{ext}$. Which implies that the methods used to derived the
formulation (\ref{eq32}) is reasonable. From the derivation we also
can know more about the relationship between discrete models and
continuous models of stochastic motion of microscopic particles.

In conclusion, we have provided an analytical formulation of
effective diffusion coefficient of the stochastic motion of
microscopic particles. The numerical comparison with the formulation
obtained in the framework of the overdamped Langevin dynamics and
Fokker-Planck equation indicates that our analytical formulation is
very accurate. Moreover, the methods used in this research can be
further used to get more results about the stochastic motion.
Through the discussion in this research, the relationship between
continuous models and discrete models has also been made clear.

\vskip 0.5cm

\noindent{\bf Acknowledgments} This work was funded by National
Natural Science Foundation of China (Grant No. 10701029).


\bibliographystyle{unsrt}

\end{document}